\documentclass{aa}  

\usepackage{newtxtext,newtxmath}
\usepackage[T1]{fontenc}
\usepackage{ae,aecompl}
\usepackage{comment}
\usepackage{aas_macros}
\usepackage[hidelinks]{hyperref}
\usepackage{graphicx}	% Including figure files
\usepackage[export]{adjustbox}

\usepackage{xcolor}

\def\gsim{\;\raise0.3ex\hbox{$>$\kern-0.75em\raise-1.1ex\hbox{$\sim$}}\;}

\def\kms{\rm ~km~s^{-1}}

\def\diff{\rm ~cm^2~s^{-1}}
\def \kms {\rm ~km~s^{-1}}

\def\ergs{\rm ~erg~s^{-1}}

\begin{document}

\title{Minimalist model of the W50/SS433 "Extended X-ray Jet": anisotropic wind with recollimation shocks}

\author{
E.~M.~Churazov \inst{1,2} 
\and
I.~I.~Khabibullin \inst{3,2,1} 
\and
A.~M.~Bykov \inst{4} 
}

\institute{
Max Planck Institute for Astrophysics, Karl-Schwarzschild-Str. 1, D-85741 Garching, Germany 
\and 
Space Research Institute (IKI), Profsoyuznaya 84/32, Moscow 117997, Russia
\and
Universitäts-Sternwarte, Fakultät für Physik, Ludwig-Maximilians-Universität München, Scheinerstr.1, 81679 München, Germany
\and
Ioffe Institute, Politekhnicheskaya st. 26, Saint Petersburg 194021, Russia
}

\abstract{W50 is a radio nebula around hyper-accreting Galactic microquasar SS~433. 
%In this letter,
Here we focus on one peculiar 
feature of W50 - a pair of so-called "extended X-ray jets" (EXJs). These "jets" have a size of $\sim20\, {\rm pc}$, a sharp inner boundary, and their spectra are well represented by a featureless X-ray continuum. We argue that EXJ could be an outcome of a powerful  {\it anisotropic} wind produced by a super-critical accretion disk. 
In the simplest version of this model, the wind itself consists of two components. The first component is a nearly isotropic outflow that subtends most of the solid angle as seen from the compact source and creates the quasi-spherical part of the W50 nebula.
The second component is a more collimated wind aligned with the binary system rotation axis (polar wind). The isotropic outflow passes through the termination shock and its increased thermal pressure creates a sequence of recollimation shocks in the polar wind,
giving it the appearance of an extended X-ray structure. In this model, the EXJ continuum spectrum is due to synchrotron emission of electrons accelerated at the shocks arising in the
polar wind. At variance with many other studies, in this model, the EXJ structures are not directly related to the highly collimated and precessing $0.26\;\!c$ baryonic jets. Instead, the EXJ and the W50's ears are produced by the part of the wind with an Eddington-level kinetic luminosity confined to a half-opening angle of 
5-10 degrees, which is not necessarily a recollimated version of the $0.26\;\!c$ jets. 
}

\titlerunning{W50 Extended X-Ray Jet}

    
\maketitle

\section{Introduction}

W50 and SS433 represent a prototypical combination of a giant radio-bright nebula and a hyper-accreting compact source that powers the nebula \citep[see, e.g.][for reviews]{1984ARA&A..22..507M,2004ASPRv..12....1F}. Many properties of W50/SS433 remain poorly understood, %though, 
partly because rare and short-living hyper-accreting black holes or neutron stars  
might operate in a very different regime than more ubiquitous and better-studied sub-Eddington sources \citep[e.g.][]{2015NatPh..11..551F}. In this regard, the stability of SS~433 properties over more than 40 years of observations \citep[e.g.][]{2021MNRAS.507L..19C} makes it a particularly appealing "laboratory" for hyper-accretor exploration \citep[e.g.][]{2006MNRAS.370..399B}.

It has been shown that the binary system in SS~433 can avoid forming a common envelope and sustain an extremely large accretion rate ($\sim$ few hundred the Eddington level) for a long time, $t_{\rm life}\sim 10^4-10^5$~yrs \citep{2017MNRAS.471.4256V,2023NewA..10302060C}. A simple estimate of the total energy that might be injected by such a source into the surrounding medium, $E_{\rm inj}\sim L_{\rm Edd} t_{\rm life}\sim 3\times 10^{50} L_{39}t_{4}\,{\rm erg}$, where  $L_{39}=L_{\rm Edd}/10^{39}\,{\rm erg\,s^{-1}}$ and $t_4=t_{\rm life}/10^{4}\,{\rm yr}$. This is comparable to the energy output of a supernova explosion that has created the compact object of the binary in the first place. As a result, such an object can strongly affect the Interstellar Medium (ISM) in the vicinity of the binary, creating and powering a giant, $\gtrsim 10$ pc, nebula  \citep[e.g.][]{1980ApJ...238..722B}. 

A key prediction of the standard accretion theory in the highly super-critical regime \citep{1973A&A....24..337S} is channeling of a large fraction of liberated gravitational energy of the accreted material into powerful trans- and mildly relativistic outflows with various degrees of axial collimation \citep[see ][for recent numerical simulations]{2022PASJ...74.1378Y}. One of the most notable features of SS~433 - the trans-relativistic baryonic jets with bulk velocity $\varv \sim 0.26\,c$ and an opening angle of less than 2 degrees - appears to be a direct confirmation of this prediction \citep[e.g. ][]{1980ApJ...238..722B,1981Ap&SS..79..387C}. These narrow jets reveal themselves in X-ray and optical spectra as pairs of blue and red-shifted lines, coming from vicinity ($\lesssim 10^{12}$ cm and $\lesssim 10^{15}$ cm for X-ray and optical lines, respectively) of the compact source \citep[][]{2004ASPRv..12....1F,2018ApJ...867L..25B, 2019A&A...624A.127W}. Modeling of these lines demonstrates that the kinetic luminosity of the jets is indeed comparable to the Eddington luminosity of a stellar-mass black hole \citep[][]{2002ApJ...564..941M,2005A&A...431..575B,2010MNRAS.402..479M,2013ApJ...775...75M,2016MNRAS.455.1414K,2019AstL...45..299M,2023A&A...669A.149F}.  The baryonic jets, however, remain invisible beyond the distance of $\sim0.1$ pc from the central source, where they manifest themselves 
%for the last time 
as a spectacular time-variable corkscrew pattern of the radio emission due to the precession of the jets' direction with a period of 163 days and half-amplitude 20 degrees \citep[e.g. ][]{1981ApJ...246L.141H,2004ApJ...616L.159B}. \footnote{The nature of the diffuse X-ray emission detected by the \textit{Chandra} Observatory from a similar location \citep[so-called "arcsec-scale X-ray emission, ][]{2002Sci...297.1673M} is less clear \citep[e.g.][]{2017AstL...43..388K}, given its irregular variability and peculiar spectral features \citep{2005MNRAS.358..860M,2008ApJ...682.1141M}. } 

Since at even large scales, from 40 to 110 pc, SS~433 is surrounded by the radio- and ${\rm H}\alpha$-bright nebula W50 with strong elongation (with 3:1 aspect ratio) along the jets' precession axis, it has long been considered that the kinetic energy of these strongly collimated jets powers/distorts the W50 nebula \citep[][]{1980ApJ...238..722B,1980MNRAS.192..731Z,1983ApJ...272...48E,1993ApJ...417..170P,2000A&A...362..780V,2008MNRAS.387..839Z,2011MNRAS.414.2838G,2014ApJ...789...79A,2015A&A...574A.143M,2017A&A...599A..77P,2021ApJ...910..149O}.

The question of whether the narrow jets power the entire nebula remains open. Indeed, in the radio, optical and X-ray images of W50, there are no direct observational signatures of the narrow jets' recollimation and/or termination \citep[like termination hot spots or a clear pattern of the decelerating flow, see, e.g., ][]{1998AJ....116.1842D,2011MNRAS.414.2838G,2017MNRAS.467.4777F,2018MNRAS.475.5360B}.  Instead, very peculiar elongated X-ray structures are seen on both sides of SS~433 \cite[e.g.][]{1983ApJ...273..688W,1994PASJ...46L.109Y,1997ApJ...483..868S,1999ApJ...512..784S,2007A&A...463..611B,2022ApJ...935..163S}. 
These features, which we further on call Extended X-ray Jets (EXJ), are separated from the binary by $\sim 20\,$pc,  aligned with the narrow jet precession axis, and have an opening angle of $\sim 20^\circ$ (i.e. twice smaller than the precession amplitude of the compact jets). They are characterized by a hard featureless X-ray spectrum that gradually steepens with distance from SS~433 \citep[e.g.][]{2007A&A...463..611B,2022ApJ...935..163S}.

The synchrotron origin of these structures is not only attractive from the spectral and energy/mass requirements points of view \citep[e.g.][]{2007A&A...463..611B}, but it also leads to verifiable predictions regarding very high energy (VHE) gamma-ray emission coming from them \citep[][]{1997ApJ...483..868S,1998NewAR..42..579A,2008MNRAS.387.1745R,2020ApJ...904..188K,2020ApJ...889..146S}. Thanks to the advent of modern VHE observatories, the 
extended emission from W50 (co-spatial with EXJ)
 has been detected at TeV energies \citep[][]{2018Natur.562...82A}. Another important implication of the synchrotron scenario is the possibility of a high degree of polarization, provided that the magnetic field is globally ordered inside these lobes. The recent observation of the eastern lobe by \textit{IXPE} revealed a high degree of X-ray polarization with the direction indicating magnetic field orientation along the main axis of the system \citep{2023arXiv231116313K}.

In this paper, we propose
a model that attempts to associate the overall W50 morphology and EXJs with an anisotropic wind from the binary system. In this model, an almost isotropic part of the wind is responsible for the quasi-spherical part of the W50 nebula, while a more collimated portion of the wind drives the "ears" of W50. In the same framework, the termination of the isotropic wind sends recollimation shocks into the collimated wind, giving rise to the EXJ features.

\section{Basic model}

We are trying to explain the following salient characteristics of the EXJs associated with W50:
\begin{itemize}
\item {\bf Sharp EXJ inner boundary} in X-rays on both sides from SS433 and the lack of bright emission in optical and radio bands.
\item {\bf Featureless EXJ X-ray spectrum}, which can (empirically) be described as either thermal bremsstrahlung with $kT\sim 4$~keV or a power law with the photon index $\Gamma\sim 2.2$ \citep{2007A&A...463..611B}. From the joint analysis of \textit{XMM-Newton} and \textit{NuSTAR} data, \cite{2022ApJ...935..163S} derived a flatter spectrum $\Gamma\sim 1.6$ for the EXJ portion closest to the SS433 and $\Gamma\sim 2.05$ for the "lenticular" region that is further away (based on the data of only \textit{XMM-Newton}).  
\end{itemize}

%2007A&A...463..611B

A usual assumption is that an outer boundary of the W50 nebula model is a forward shock associated with the supernova explosion and a jet/outflow \citep[e.g.][]{2011MNRAS.414.2838G}. Since the inner boundary of EXJ is closer to the binary than the quasi-spherical part of W50, it is not easy to associate it with any characteristic radius. We assume instead that the entire nebular is powered by a wind produced by SS~433. The wind has an additional characteristic radius - a termination shock, which we want to associate with the inner boundary of EXJ. A wind from a companion star was considered by \cite{1983MNRAS.205..471K}, while \cite{2014ApJ...789...79A} assumed that the wind is produced by the super-Eddington accretion onto the compact object, which is similar to the assumption adopted here.  Indeed, for the wind mechanical power of $\sim 10^{39}\,{\rm erg\,s^{-1}}$, the total energy produced by the wind over the estimated age of $\sim 6\times 10^{4}\,{\rm yr}$, is $\sim 2\times 10^{51}\,{\rm erg}$, of the same order or even larger than the kinetic energy of SN ejecta. Here, we ignore the contribution of the SN energy input, although such a scenario is often assumed.  What matters is that at the present epoch, the powerful wind is present and it has already been operating for a sufficiently long time. We briefly discuss such a case in Appendix~\ref{app:snr}.

A steady isotropic wind model \citep{1985Natur.317...44C} is characterized by a set of four parameters (assuming that the wind is supersonic): $L_{\rm w}$, $\varv_{\rm w}$, $t_{\rm age}$, and $n_0$, which are the wind kinetic power, wind velocity, age of the system, and the density of the interstellar medium (ISM), respectively. The latter is assumed to be homogeneous and uniform. In the frame of this model, one can estimate these four parameters from several basic observables, such as, e.g., the radius of the nebula and the gas temperature downstream of the shock driven by the wind. 

\begin{figure}
\centering
\includegraphics[angle=0,trim=1cm 5.5cm 1cm 2.5cm,width=0.95\columnwidth]{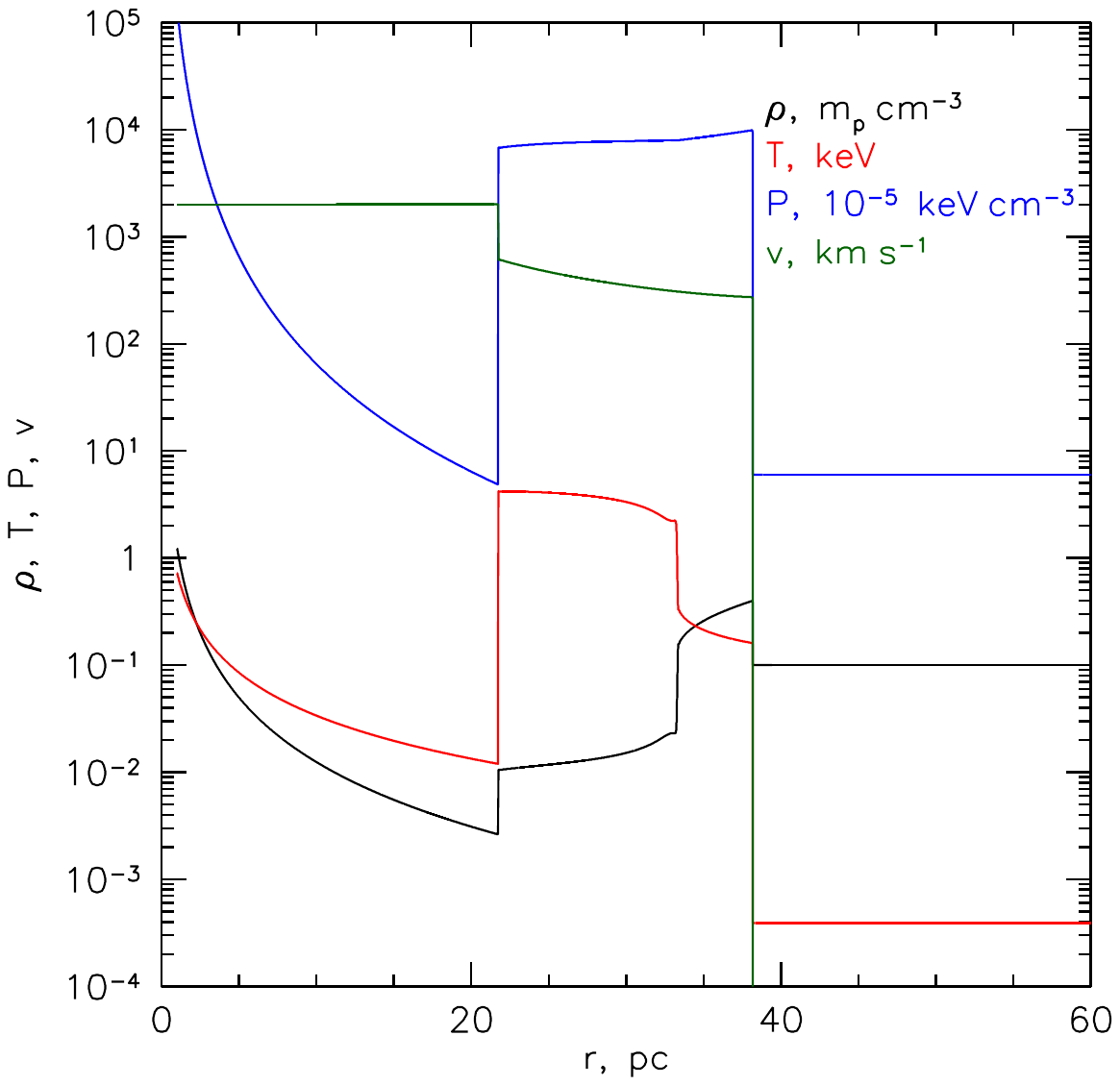}
\caption{An example of a spherically symmetric wind model for $\rho_{\rm ISM}=0.1~{\rm m_p\, cm^{-3}}$, $L_{\rm w}\sim 10^{39}~{\rm erg~s^{-1}}$, $\varv_{\rm w}=2\,10^3\;{\rm km\,s^{-1}}$ after $t_{\rm age}\sim 6\,10^{4}\,{\rm yr}$. 
%$\dot{M_W\sim 1.3\,10^{-4}}$ 
With these parameters, the densest region is the shell of the shock-compressed ISM on the downstream side of the forward shock, which should dominate the X-ray emission of the nebula. The density of the shocked wind material is too low to produce detectable X-ray emission. 
}
\label{fig:wind_def}
\end{figure}

A typical radial structure of the wind expanding into a uniform medium is shown in Fig.\ref{fig:wind_def}. It consists of a forward shock advancing into the ISM, a wind termination shock, and a contact discontinuity, separating the shocked ISM from the shocked wind material. We associate the quasi-spherical part of the W50 nebula with the forward shock. 
Assuming that the forward shock is strong and adopting 
the 
downstream temperature of the gas of $T_{s}\sim 0.2\,{\rm keV}$ \citep[e.g][]{2007A&A...463..611B,2022ApJ...935..163S}, the 
%forward 
shock velocity is
\begin{equation}
\varv_s=\left (\frac{16}{3} \frac{kT_{s}}{\mu m_p}\right)^{1/2}\approx 410 \left(\frac{kT_{s}}{0.2~{\rm keV}}\right)^{1/2}~{\rm km\,s^{-1}} .
\end{equation}
The estimate of the age of the system for a supersonic expansion, i.e. $R_s\propto t^{3/5}$ is then 
\begin{equation}
t_{\rm age}\approx \frac{3}{5}\frac{R_s}{\varv_s}= 5\,10^4 \left (\frac{R_s}{38\,{\rm pc}}\right ) \left (\frac{\varv_s}{410\,{\rm km\,s^{-1}}} \right )^{-1} \; {\rm yr},
\end{equation}
while the power of the wind/outflow is
\begin{eqnarray}
L_{\rm w}\approx \frac{\frac{4}{3}\pi R^3_s \frac{9}{4}n_0 \mu m_p \varv_s^2}{t_{\rm age}}=
\\ 7\times 10^{38} \left (\frac{n_0}{0.1\,{\rm cm^{-3}}}\right ) \left (\frac{R_s}{38\,{\rm pc}}\right )^2 \left(\frac{kT_{s}}{0.2~{\rm keV}}\right)^{3/2} {\rm erg\,s^{-1}},
\end{eqnarray}
where $n_0$ is the particle number density of the unshocked ISM, $\mu\approx 0.61$ is the mean atomic weight per particle for the fully ionized gas with a standard cosmic composition. Here we assume that the forward shock is strong and the energy density is constant across the entire volume subtended by the shock. Among the three observables used above, namely $R_s$, $T_{s}$ and $n_0$, only
$R_s$ can be easily measured directly from X-ray or radio images, while $T_s$ and $n_0$ can be robustly determined from X-ray spectra only if the shock-heated ISM has reached collisional ionization equilibrium (and $T_e\approx T_i$), which is not guaranteed a priori. So we use some fiducial values for $T_s$ and $n_0$, but acknowledge that their values (except for $R_s$) are uncertain by a factor of a few. For instance, the density $n_0$ can be lower, while the post temperature $T_s$ can be higher \citep[see, e.g., a non-equilibrium ionization model in][]{2007A&A...463..611B}, making the estimate power uncertain in either direction. In particular, their best-fitting estimate of ionization time $n_et\sim 3\times 10^{10}\,{\rm s\,cm^{-3}} $ for $n_0\sim 0.1\,{\rm cm^{-3}}$ (for ionized plasma with solar abundance of elements, $n_e\approx 0.52\, n_0$) translates into $t\sim 2\times 10^4\,{\rm yr}$. It is, therefore, plausible that the actual density of the ambient medium is a factor of a few lower.

The velocity of the wind $\varv_{\rm w}$ could be estimated if the radius of the wind termination shock $R_T$ is known. The pressure is approximately constant between the forward shock position $R_s$ and $R_T$. The latter is determined from the balance of the ram pressure of the wind $P_{\rm w}$ and the postshock pressure $P_{s}$, i.e.
\begin{equation}
P_{\rm w}=\rho_{\rm W} \varv_{\rm w}^2=\frac{L_{\rm w}}{4\pi R^2_T \varv_{\rm w}}\sim P_{s}=4\, n_0 \frac{3}{16} \mu  m_p \varv_s^2,
\end{equation}
where $\rho_{\rm w}(R)\propto R^{-2}$ and $\varv_w$ are the wind density and velocity, respectively, and the factor of $4$ stands for the compression ratio at the forward shock, which is assumed to be strong.
The above relation (dropping factors of order unity) implies that the radii of the forward and termination shocks are related via the ratio of the shock and wind velocities as
\begin{equation}
R_{\rm T}\approx R_s \left ( \frac{\varv_s}{\varv_{\rm w}}\right )^{1/2}.
\end{equation}
%Isotropic outflow, energetics, termination shock, Koenigle..

%------------------------
%-------------------------

The isotropic wind expanding into a homogeneous medium naturally produces a spherically symmetric forward shock, a contact discontinuity, and a termination shock (see Fig.\ref{fig:wind_def}).  The density of the shocked wind might be low enough so that its thermal X-ray emission is too faint to be directly observed. Indeed, from Fig.~\ref{fig:wind_def} that corresponds to the \cite{1985Natur.317...44C} steady wind solution, it follows that, for the adopted parameters, the density downstream of the termination shock is a factor of $\approx 40$ smaller than that at the forward shock. Given that the X-ray emissivity scales as the square of the gas density, the expected surface brightness of the termination shock region at energies below a few keV is a factor of about $10^3$ smaller than for the forward shock.
Also, a termination shock of the isotropic wind is not necessarily an efficient particle accelerator \citep[e.g.][]{2004A&A...424..747P}.  Therefore, in this model, it is not surprising that we see neither radio nor X-ray emissions from this shock. 

The W50 nebula is, however, elongated in the direction perpendicular to the orbital plane. This suggests that the wind is anisotropic and, in particular, more powerful (per unit solid angle) along this direction. To explain this morphology, a combination of expanding spherical SN shock and a jet/outflow is often used \citep[e.g.][]{2011MNRAS.414.2838G}. Here, we model this configuration as a two-component wind. One component is a nearly isotropic wind considered above (hereafter "i-wind"), while the second component is a collimated polar wind (hereafter "p-wind") along the orbital momentum of the binary. The main difference from the other models is the presence of a termination shock of the isotropic wind that recollimates the p-wind. We reiterate here, that the postulated "p-wind" is not the highly collimated subrelativistic jet, but a part of the wind associated with inner regions of the accretion flow that subtends a much larger solid angle than the narrow jets.  

Propagation of jets in astrophysical conditions has been extensively studied both theoretically and numerically \citep[e.g.][]{1998MNRAS.297.1087K,2012MNRAS.427.3196K,2023JPlPh..89e9101P}. For example, for radio sources associated with Active Galactic Nuclei (AGN), a canonical model identifies several characteristic length scales that determine jet propagation. These include a scale where the jet becomes underdense with respect to the ambient gas, a scale where the ambient pressure can counterbalance the side expansion of a conical jet and a scale where the ram pressure of the jet is equal to the ambient pressure. These scales determine whether the jet is forming a cocoon or is recollimated by external pressure, and the position of the terminal shock. In the model discussed here, the terminal shock of the isotropic wind introduces yet another length scale, which modifies the behavior of the p-wind, which can itself be treated as a jet. In addition, the propagation of the p-wind is happening in the expanding i-wind, i.e. the densities and pressures of the two flows have similar dependencies on the radius (before reaching the i-wind termination shock). Therefore, the relation between densities and pressures is set at the base of the two flows and is further modified at the position of the i-wind terminal shock (see Appendix~\ref{app:winds}).

A jet/wind interaction has also been studied in the context of massive binary systems \citep[][]{2022MNRAS.510.3479B,2022A&A...661A.117L}. There, a jet/outflow produced by an accreting black hole collides with the wind coming from a massive star. In the configuration considered here, it is assumed that accretion on SS433 proceeds via the Roche Lobe Overflow (RLOF) \citep[][]{2017MNRAS.471.4256V,2021MNRAS.507L..19C} rather than via stellar wind. In this case, the i-wind and p-winds both come from the accretion disk and move (radially) in the same direction rather than collide.   

Throughout the paper, we assume that the termination radius for the p-wind is larger than for the nearly isotropic component (i.e., the dynamic pressure $\rho \varv^2$ at a given radius is larger for the p-wind). This means that the material in the collimated component will "see" a sudden increase of the ambient pressure when crossing the termination shock of the i-wind. In aerodynamics, this corresponds to the case of an "over-expanded jet", when the internal jet pressure at the edge of the nozzle is lower than the ambient pressure. This large pressure will send a recollimation shock into the p-wind, compressing and heating it. Therefore, an additional characteristic radius appears in the problem, which we associate with the inner boundary of EXJs well inside the forward shock. This is the essence of the proposed model. 

\begin{figure*}
\centering
\includegraphics[angle=0,trim=3cm 3cm 4cm 5cm,width=1.6\columnwidth]{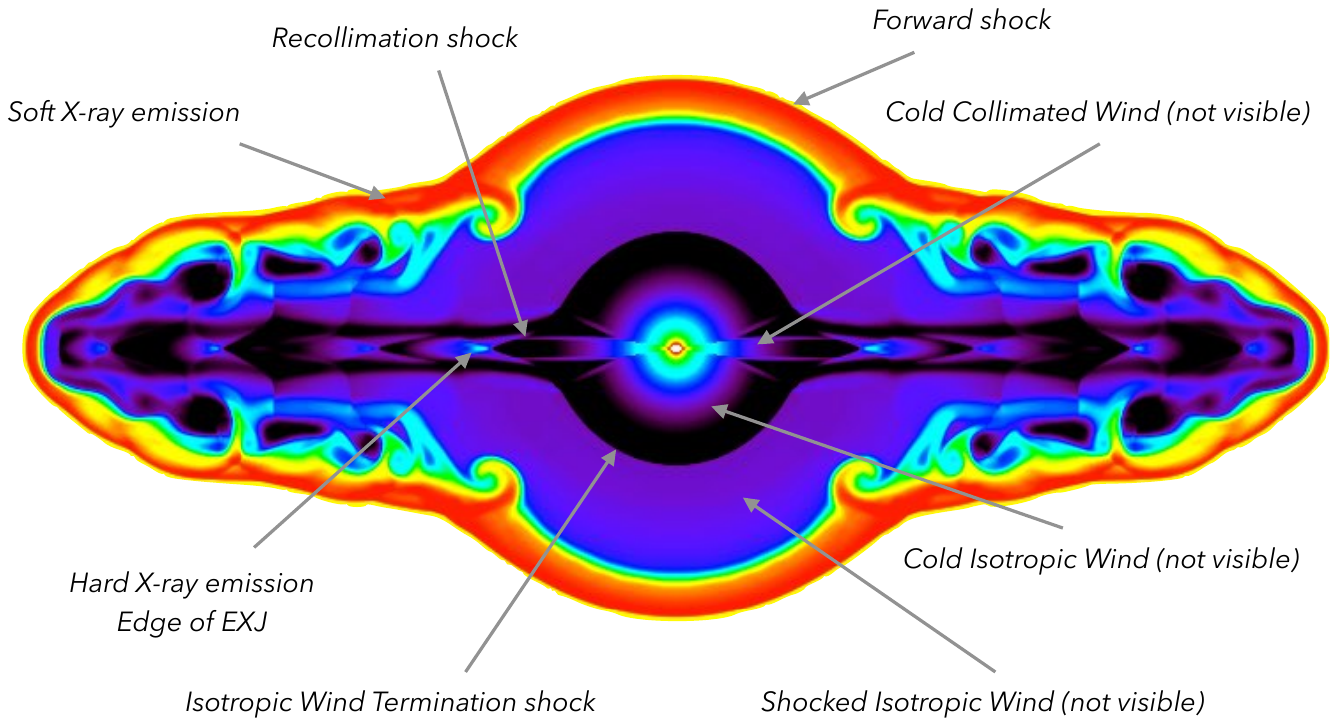}
\caption{Sketch of the W50 model as a combination of isotropic and polar winds. The central region represents a combination of an isotropic wind and a more collimated "polar" wind aligned with the binary system orbital axis. The isotropic wind passes through its termination shock and increased pressure and density recollimates the polar wind. Resulting recollimation shocks accelerate particles and give rise to the synchrotron emission of the EXJs.}
\label{fig:sketch}
\end{figure*}

To illustrate some basic features of the model, we run a set of 2D simulations (axisymmetric cylindrical symmetry) using the \texttt{PLUTO} code \citep{2007ApJS..170..228M}. The details of the simulations and the choice of parameters are described in Appendix~\ref{app:winds}. A labeled sketch 
(density slice) is shown in Fig.~\ref{fig:sketch} for $t_{\rm age}\sim 6\times 10^4\,{\rm yr}$ and the i-wind kinetic luminosity of $10^{39}\,{\rm erg\,s^{-1}}$. In supercritical accretion disks, a massive outflow starts at the "spherisation" radius $\displaystyle R_{\rm sp}\approx 10 \, \dot{m_0} \frac{GM}{c^2}$  \citep{1973A&A....24..337S,2007MNRAS.377.1187P}, where $M$ is the mass of the compact object and $\dot{m_0}$ is the accretion rate in Eddington units. The corresponding escape velocity near $R_{\rm sp}$ is $\displaystyle \varv_{esc}\sim \sqrt{\frac{GM}{R_{\rm sp}}}\sim c/\sqrt{10\,\dot{m_0}}\sim 3000\,{\rm km\,s^{-1}}$ for $\dot{m_0}=10^3$. In Fig.~\ref{fig:sketch} we set $\varv_i=2000 \,{\rm km\,s^{-1}}$ and in the Appendix~\ref{app:winds} we consider cases with $\varv_i=2000$ and $4000\,{\rm km\,s^{-1}}$.

The outflow continues at smaller radii of the supercritical disk. In our simplified model, we attribute these "inner" outflows to the p-wind component. In Fig.~\ref{fig:sketch}, the p-wind component is initially confined to a cone with a half-opening angle of $\sim 9$ degrees and its density and velocity are both a factor of 2-3 higher than that in the i-wind. The kinetic luminosity of the p-wind is $\sim 14$\% of the i-wind.  
Since we are interested in the global picture of the nebula out to 100~pc, the winds are initiated within a sphere with radius $8\,{\rm pc}$. The radius of the p-wind cross-section at this sphere is $\sim 1.3\,{\rm pc}$.  We note here that the model is flexible enough to broadly reproduce the overall morphology of the nebular for different sets of parameters, e.g. a faster p-wind or a much denser and slower p-wind (see Appendix~\ref{app:winds}). However, these models share several common features, which we highlight in Fig.~\ref{fig:sketch}. In this figure, the termination shock of the isotropic wind corresponds to the transition from the lowest density region (isotropic wind component just before the shock) to the 4 times larger density (downstream from the termination shock). The collimated outflow goes through a series of recollimation shocks, where the dense regions are interleaved with the lower-density patches. While the first dense region $R_{d}$ does not occur exactly at the position of the isotropic wind termination shock, they are related, so that $R_{d}\sim O(1)\times R_T$, where the pre-factor sensitively depends on the p-wind opening angle and relative densities and velocities of the winds. The choice of the parameters for this simulation is rather arbitrary and other combinations can lead to a similar morphology when the density or velocity ratios between the i-wind and p-wind are changed together with the characteristic opening angle of the p-wind.  However, the recollimation shocks are always present when the i-wind termination shock is located at a smaller radius than that of the p-wind.

The above model completely neglects the contribution of the supernova explosion to W50. The primary motivation for this assumption is a mismatch between the timescale of stellar evolution (millions of years) and the duration of the period of extremely high accretion of the binary during the RLOF phase ($\sim10^5\,{\rm yr}$) \citep[see, e.g.][]{2017MNRAS.471.4256V,2023NewA..10302060C}. If the former determines the onset of the ROLF after the supernova explosion, it is plausible that at present time the traces of the supernova remnant are largely gone. Given the peculiar nature of SS433/W50, one might imagine that this binary system has a similarly peculiar history and the supernova explosion took place some $\sim 10^5\,{\rm yr}$ ago \citep[e.g.][]{2011MNRAS.414.2838G,2017A&A...599A..77P,2018A&A...617A..29B}. We consider this possibility in Appendix~\ref{app:snr}. In brief, adding $\sim 10^{51}\,{\rm erg}$ in the form of the SN ejecta can certainly contribute to nebular size and shape, but if the wind is established a sufficiently long time ago, the position of the wind termination shock will be the same as shown in Fig.~\ref{fig:snr} and the
overall morphology may remain similar to the one depicted in  Fig.~\ref{fig:sketch}.

%------------------------
%-------------------------

\section{Discussion}
Several models have been proposed to explain the spectrum of the EXJs \citep[e.g.][and references therein]{2022ApJ...935..163S}, which attribute it to synchrotron emission. However, the origin of the EXJs themselves, which "appear in the middle of nowhere" is not easily explained. 

Previously, we have considered an exotic scenario for the formation of EXJ,  when a very fast (sub-relativistic) outflow recombines and impacts the ISM in the form of a beam of neutral atoms \citep{2020MNRAS.495L..51C}. In that model, {\it neutral} particles penetrate the ISM, get ionized, and deposit their energy in a long and narrow "channel" along the direction of the beam. The heated ISM expands sideways, reducing the gas density and letting neutral particles propagate further without being ionized.  This (i) leads to a fast propagation of the heating front and (ii) gives rise to a system of diverging and converging shocks perpendicular to the beam. This analysis was not specifically tied to W50/SS433, except for the assumption that neutral particles are moving with the velocity $\sim 0.25\,c$. In the neutral beam model, the EXJ structure forms where the beam enters the dense medium. Therefore, to explain the inner boundary of the EXJ one has to postulate the presence of dense gas at this location.

The model, discussed in the previous section is different. It does not rely on the SS433 trans-relativistic baryonic jets. Instead, it associates EXJ with the recollimation shocks in the anisotropic wind coming from the SS433 binary system. In this model, the isotropic part of the wind reaches the termination shock earlier than the polar wind. This translates into a condition that momentum flux per unit solid angle is larger for the p-wind 
\begin{equation}
\rho_p \varv_p^2 > \rho_i \varv_i^2,
\end{equation}
which is equivalent to the condition
\begin{equation}
\frac{L_p}{\Omega_p \varv_p} > \frac{L_i}{\Omega_i\varv_i},
\end{equation}
where $L_p$ and $L_i$ are the kinetic luminosities of the p- and i-winds, respectively, and $\Omega_p$ and $\Omega_i$ are the corresponding solid angles subtended by the winds.
In the illustrative example shown in Fig.~\ref{fig:sketch}, $\rho_p\varv_p^2\approx 13 \rho_i \varv_i^2$, i.e. the above condition is satisfied.

Downstream of the termination shock in the isotropic wind, the gas becomes hot and 4 times denser. As mentioned above, the gas density is too low to give rise to observable thermal (bremsstrahlung) emission. The termination shock of the i-wind resembles the case of an almost
perpendicular shock (upstream magnetic field is along the shock front)
similar to stellar winds around massive stars. No signature of synchrotron X-ray emission was ever reported from the termination shocks of fast radiative-driven winds of young massive stars.  Particle acceleration there is likely much less efficient than that in supernova shocks of similar velocity \citep[e.g.][]{2001SSRv...99..329D,2004A&A...424..747P}.  Cosmic ray acceleration in the vicinity of the solar wind termination shock (of speed above 400 $\kms$) directly measured by the Voyagers is also rather modest \citep[][]{2022SSRv..218...42R}. Recent modeling of CR acceleration by stellar wind bubbles of young massive stars by \citet{2024arXiv240318484M} demonstrated that the systems can mostly re-accelerate pre-existing cosmic ray particles at the bubble forward shocks rather than accelerate CRs at the wind termination shocks.    
Thus the i-wind termination shock can have a low
efficiency of particle acceleration. On the contrary, the system of
recollimation shocks formed due to i- and p-winds collision can have a much more complicated structure of
magnetic fields, involving both quasi-parallel and quasi-perpendicular
region. Here, we simply assume that the acceleration of particles in the
recollimated p-wind proceeds similarly to quasi-parallel shocks with a compression factor of 4 (might be larger if CR escape effects are accounted
for) and the characteristic shock velocity $\sim \varv_p$.

The gas in the p-wind experiences oblique recollimation shocks and Mach disks driven by the isotropic wind, it heats up and goes through a sequence of compression and rarefactions. Similarly to the i-wind, the shocked p-wind density is far too low to produce appreciable thermal emission. 
%In principle, one can assume that the density ratio relative to the isotropic wind is high enough so that at the position of the recollimation shock the emission measure of the shocked gas is sufficiently large. But in this case, the temperature of the shocked gas will be too low to explain the spectral properties of EXJs. 
We, therefore, rely on the possibility that recollimation shocks are effective in accelerating particles and the EXJ emission is non-thermal  \cite[as suggested before, see e.g. ][ and references therein]{2022ApJ...935..163S}. Indeed, recent detection of X-ray polarization from EXJ \cite{2023arXiv231116313K} with \textit{IXPE}, shows that X-ray emission has a synchrotron origin, while the magnetic field is largely along the Eastern EXJ. The acceleration could be due to shocks \citep[i.e. Diffusive Shock Acceleration or DSA][]{1977DoSSR.234.1306K,1978MNRAS.182..147B} or due to a shear flow \citep[][]{1981ICRC....3..506B,2002ApJ...578..763S,2019ApJ...886L..26R} that naturally appears in the configuration considered here. In both cases, it is plausible that the acceleration is not confined only to a localized region at the first recollimation shock but is distributed along the EXJ.

The hard non-thermal X-ray emission up to 30 keV from a "knotty" region within the Eastern EXJ was detected with the \textit{NuSTAR} telescope and attributed to synchrotron radiation of very high energy electrons with Lorentz factors above 10$^7$ \citep{2022ApJ...935..163S,2020ApJ...889..146S}. These electrons have to be (re)accelerated locally. An ensemble of the recollimation shocks that have to be present after the collision of the p-wind with the termination shock of the i-wind is a plausible site of particle acceleration and magnetic field amplification. Non-thermal X-ray emission observed as thin filaments at the edge of the shell-type young supernova remnants is almost certainly produced by synchrotron emission of TeV regime electrons radiating in highly amplified magnetic field \citep[][]{2008ARA&A..46...89R, 2012SSRv..173..369H,2023ApJ...945...52F}. Particle acceleration by strong shocks of velocity well above a few thousand $\kms$ in young supernova remnants is likely accompanied by strong amplification of magnetic fluctuations resulting in Bohm-like diffusion of accelerated particles in a wide interval of particle energies \citep[see e.g.][]{2014ApJ...789..137B, 2014ApJ...794...47C}. In the case of Bohm diffusion, the acceleration time $t_a$ of a particle with Lorentz factor $\gamma$ by a shock of velocity $\varv_{p}$ scales with the magnetic field strength $B$ as $t_a \propto \gamma \varv_{p}^{-2} \eta B^{-1}$. Here the diffusion coefficient of a relativistic particle of mass $m$ is $D = \eta c^2 m\gamma /eB$, where the dimensionless factor $\eta \ge$ 1.  The maximal Lorentz factor achievable by the electrons and positrons accelerated by the shock is limited by the synchrotron losses with $\gamma_m \propto \varv_{p} B^{-1/2} \eta^{-1/2}$. Therefore, in a single zone model (where particle acceleration and the synchrotron radiation are spatially coincident) the maximal energy of the synchrotron photon $E_m$ depends mostly on the shock velocity  $E_m \propto \varv_{p}^2/\eta$. To get $E_m \approx$ 30 keV one needs $\varv_{p} \sim 20,000\,\eta^{1/2} \kms$. The energy flux of the amplified magnetic field in the shock downstream may reach 0.1 of the shock ram pressure \citep{2014ApJ...789..137B}, which is consistent with observations of young supernova remnants \citep{2012SSRv..173..369H}. In this case, the magnetic field just in the downstream of the recollimation shock can be estimated from the relation $B^2/8\pi \approx 0.4 \frac{L_p}{\Omega_i R_T^2 \varv_{p}}$. The magnitudes of the fluctuating magnetic field amplified by cosmic-ray-driven instabilities may reach $\sim$ 100 $\mu$G. Then the accelerated electrons radiating 30 keV synchrotron photons should have energies $\sim$ 100 TeV. The spectra of electrons accelerated by the DSA Fermi mechanism are consistent with those derived from hard X-ray observations \citep{2022ApJ...935..163S,2022PASJ...74.1143K}.

When this paper was under revision, new results from H.E.S.S. observations of W50/SS433 became available \citep{2024Sci...383..402H}. From a model fitting of spatial profiles of the gamma-ray emission along the jet axis, the starting velocity 
for both the eastern and western jets was found to be $0.083 (\pm 0.026{\,\rm stat} \pm
0.010{\,\rm syst})\,c\approx 25000\,{\rm km\,s^{-1}}$.
Accelerated electrons are continuously injected with a power law spectrum of index 2 and cut-off energy > 200 TeV at the outer jet base in the 1D propagation model discussed by \citet{2024Sci...383..402H}. The leptons then travel along the jet producing the observed radiation. They considered two models of the external jets propagation and estimated the diffusion coefficients of relativistic 100 TeV particles to be $(2.3 \pm 1.4) 10^{28} \diff$ assuming a decelerating jet and ($4.7 \pm 4.1) 10^{27} \diff$ for the jet of a constant velocity. The model assumed a homogeneous magnetic field along the jets of a magnitude close to 20 $\mu$G. In our model discussed above the maximal RMS amplitude of a fluctuating magnetic field can be a few times higher and reach $\sim$ 100 $\mu$G in the shock recollimation region. The turbulent fluctuating field would decay outside the acceleration region where CR-driven instabilities amplify the field. The energies of the synchrotron X-ray emitting electrons can be $\sim$ 100 TeV.

Protons can be accelerated by DSA mechanism with Bohm's diffusion in the shock vicinity to the maximal energies $E_{pm} \gsim  {\rm PeV}\, \varv_{s01}\, R_{\rm pc}\, B_{100}\, \eta^{-1}$,  where $\varv_{s01}$ is the shock velocity measured in units of 0.1 c, $R_{\rm pc}$ is the size of acceleration region measured in pcs and $B_{100}$ is the magnitude of the turbulent magnetic field in the acceleration region measured in units of 100 $\mu$G.

The protons accelerated above 100 TeV in the recollimation region will leave the accelerator and can contribute to multi-TeV regime emission detected by HAWC \citep{2018Natur.562...82A} and LHAASO \citep{2023arXiv230517030C}. Assuming the lifetime of EXJ $\sim 10^4-10^5\,{\rm yr}$, these protons will radiate about tenths of a percent of their power due to inelastic hadronic collisions in $\sim$ 30 pc vicinity of the knot 
%filled with a plasma of number density $%n$ (measured in $\cmc$) 
provided that the diffusion coefficient of the high energy protons is $\sim$ 10$^{27} \diff$. The value is generally consistent with that estimated in the vicinity of typical TeV-halos \citep[see e.g.][]{2017Sci...358..911A,2022A&A...666A...7M,PhysRevD.109.043041}. We note here that the studied TeV-halos are produced by the rotation power of pulsars (such as Geminga) with much lower kinetic/magnetic luminosity than that of W50/SS433 which is powered by hyper-accretion. In this model, the power of the accelerated very high energy nuclei can reach $\sim$ 10$^{37} \ergs$. Eventually, these nuclei should leave W50/SS433 and contribute to the galactic cosmic ray population.  

To make the simplified quantitative estimations of particle acceleration and high energy radiation given above we limit ourselves to the DSA model description.
Nonlinear DSA models constructed for single plane shocks account for cosmic ray driven instabilities \citep[see e.g.][]{2012SSRv..173..491S} that may highly amplify ambient magnetic field fluctuations and thus govern the high energy particle transport. The recollimation of p-wind is accompanied by the formation of a system of oblique shocks and the description of particle acceleration in the wind recollimation region is more complex than the DSA model used above. Namely, the spectra formation at the highest energy end can be affected by CR interaction with the shear flows \citep[see e,g.][]{2019ApJ...886L..26R} which is associated with the p-wind recollimation region.

Accurate modeling of particle spectra would require both high-resolution simulations of the recollimation region for different possible magnetization of p-wind and kinetic modeling of high-energy particle transport in the region.    

The simulations shown in Figs.~\ref{fig:sketch} and \ref{fig:jets} are not intended as accurate models of the S433/W50 system. They are used solely to illustrate a range of possible morphologies and to argue that in this hyperaccreting binary, an "extended jet-like" feature might arise naturally even in the absence of an SNR and the famous narrow $0.26\,c$ jet,  which are at focus in many other studies. We also do not discuss the role of disk precession and a more realistic wind configuration. We leave these issues and detailed high-resolution magneto-hydrodynamic simulations of the setup discussed here for future studies.

\section{Conclusions}

We outline a possible model of Extended X-ray Jets (EXJs) in the W50 Nebula. At variance with other popular models, we do not associate these structures with the well-collimated $0.26\,c$ baryonic jets produced by the hyperaccreting binary SS433. It is proposed instead that a powerful anisotropic wind from the binary that was active for a few tens of kyrs is responsible for these structures (and overall W50 morphology as well). In the simplest version of this model, the wind consists of two components: an almost isotropic wind that subtends most of the solid angle and a more collimated "polar" wind. Both wind components are cold and "invisible" 
%\ik{m.b. non-interacting?} 
until the isotropic wind reaches its terminal shock. 
%Beyond this point, a series of recollimation shocks generate relativistic electrons that power a featureless (synchrotron) X-ray spectrum, which is observed as EXJs 
Beyond this point, conical recollimation shock is launched into the polar wind component, which is capable of accelerating relativistic electrons powering spectrally featureless and polarized synchrotron radiation visible in X-rays as EXJs.

The inner edges of the EXJs correspond approximately to the location of the first recollimation shock "knot". Plausibly, as suggested earlier, the TeV emission is produced by the same electrons due to IC scatterings of CMB photons. This model also associates the overall shape of the W50 Nebula (a sphere plus two "ears") with the same anisotropic wind, namely the forward shock of the wind propagating into the undisturbed ISM. A particular set of parameters used here for illustration is loosely constrained and likely does not match exactly those in W50. However, we believe that recollimation of the anisotropic wind offers an attractive explanation of the peculiar properties of the Extended X-ray jets in SS433/W50.  It is plausible that other hyperaccreting systems can give rise to similar structures as long as the wind has a sufficiently strong angular modulation pattern. The model also stipulates that W50 EXJs are powerful sources of very high-energy cosmic ray protons, which are accelerated within the same converging flow region.

\clearpage

\section*{Acknowledgements}
We thank our referee for many helpful and constructive comments. E.C. and I.K. are grateful to  Selma E. de Mink and Stephen Justham for a useful discussion on the binary evolution of SS433-like systems. We thank Alexander Panferov for useful comments. I.~K. acknowledges support by the COMPLEX project from the European Research Council (ERC) under the European Union’s Horizon 2020 research and innovation program grant agreement ERC-2019-AdG 882679. A.~B. was supported by the baseline project FFUG-2024-0002 at Ioffe Institute. Some modeling was made at the Supercomputing Center of the Peter the Great Saint-Petersburg Polytechnic University.
\bibliographystyle{aa}
%\bibliography{references} % if your bibtex file is called example.bib
\bibliography{ref} % if your bibtex file is called example.bib

%%%%%%%%%%%%%%%%%%%%%%%%%%%%%%%%%%%%%%%%%%%%%%%%%%

%%%%%%%%%%%%%%%%%%%% REFERENCES %%%%%%%%%%%%%%%%%%

% The best way to enter references is to use BibTeX:

%\bibliographystyle{mnras}
%\bibliography{example} % if your bibtex file is called example.bib

%%%%%%%%%%%%%%%%%%%%%%%%%%%%%%%%%%%%%%%%%%%%%%%%%%

%%%%%%%%%%%%%%%%% APPENDICES %%%%%%%%%%%%%%%%%%%%%

\begin{appendix}

\section{W50 morphology for different wind parameters}
\label{app:winds}
\begin{figure*}
\centering
\includegraphics[angle=0,trim=0cm 0cm 0cm 0cm,width=0.8\columnwidth]{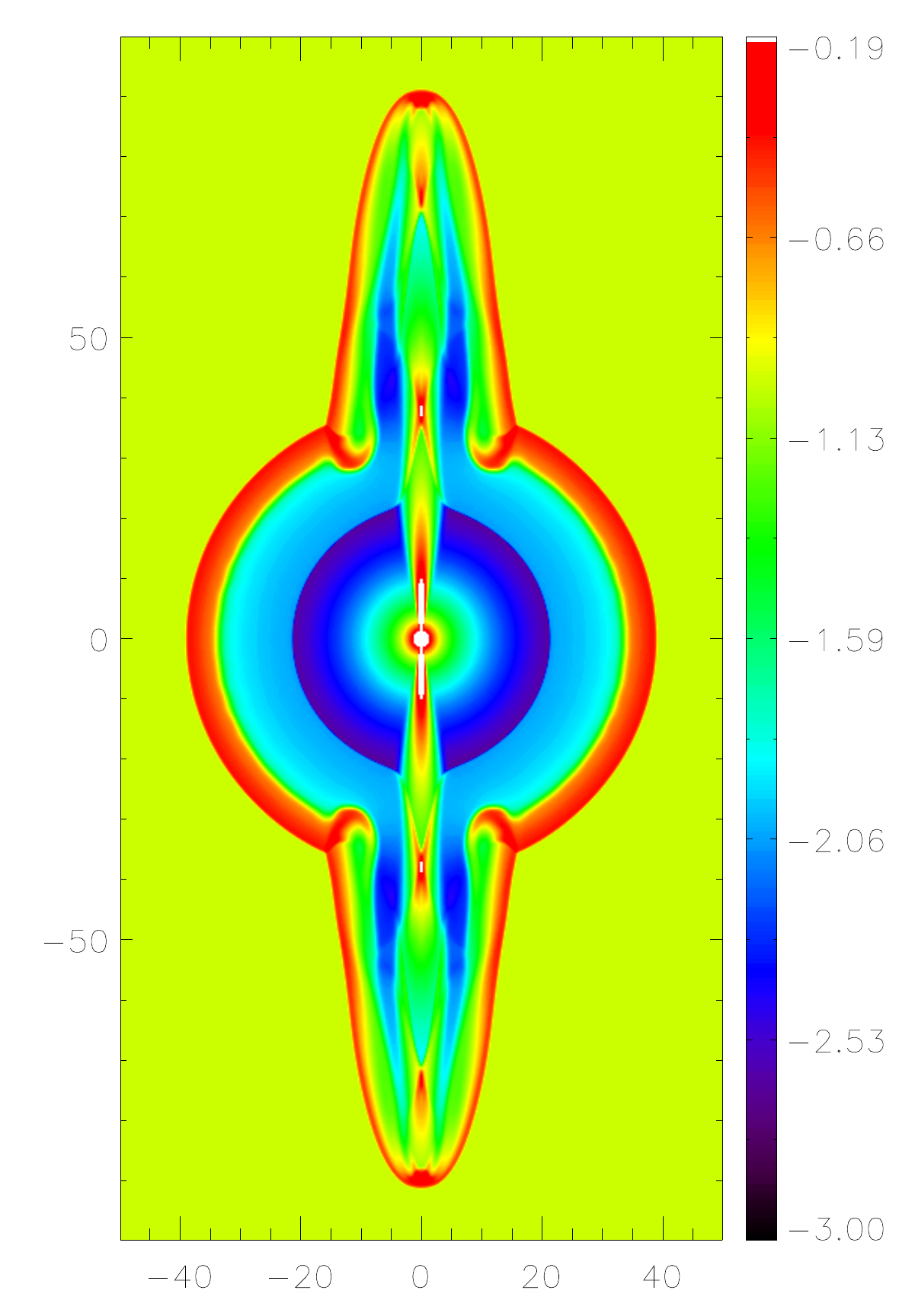}
\includegraphics[angle=0,trim=0cm 0cm 0cm 0cm,width=0.8\columnwidth]{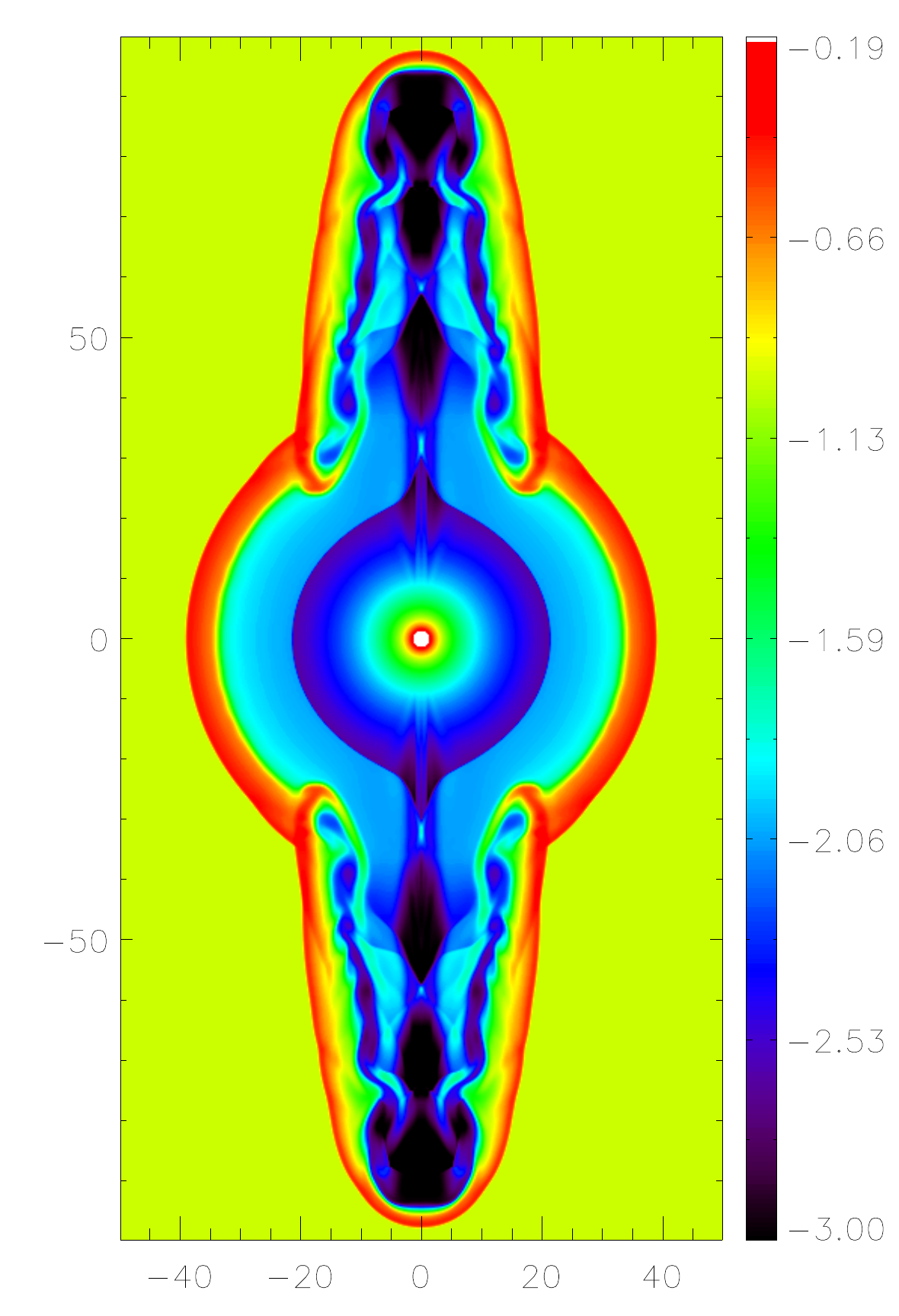}
\includegraphics[angle=0,trim=0cm 0cm 0cm 0cm,width=0.8\columnwidth]{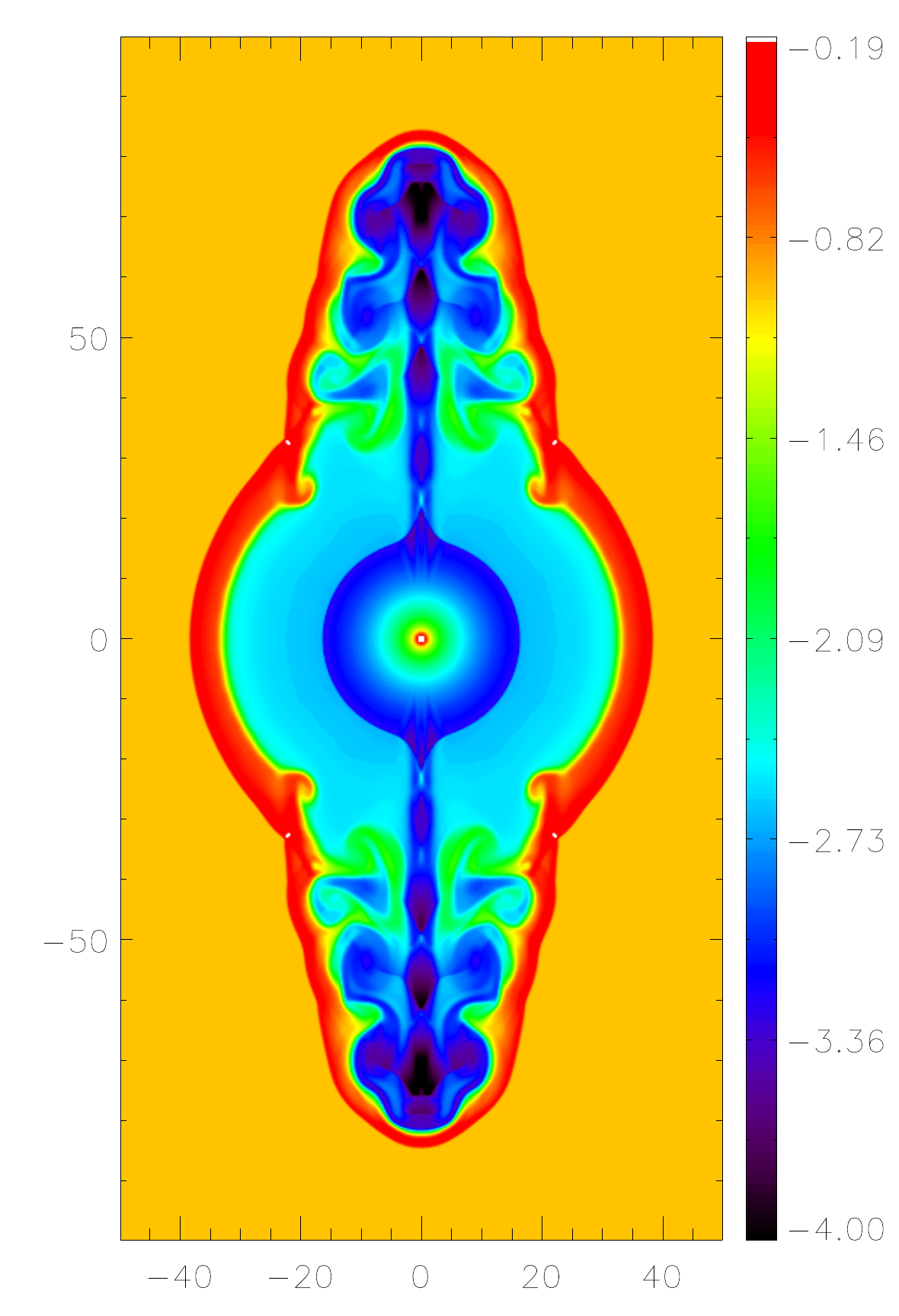}
\includegraphics[angle=0,trim=0cm 0cm 0cm 0cm,width=0.8\columnwidth]{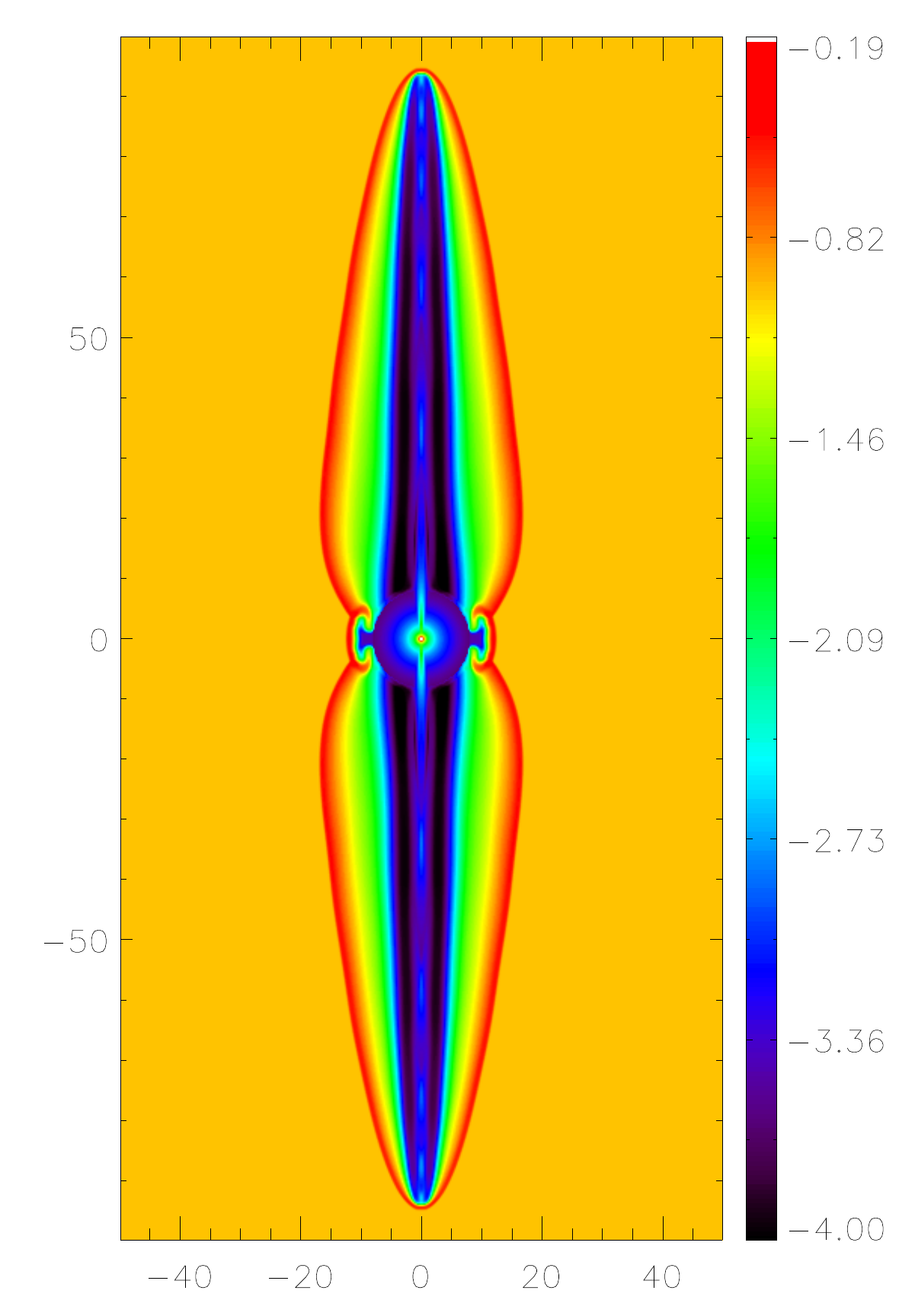}
\caption{ Density slices for different wind configurations. The color corresponds to $\log_{10}(\rho)$, where $\rho$ is in units of $m_p/{\rm cm^3}$ and the color-coding is different for the top and bottom rows. The top two panels use the same parameters of the i-wind: $L_i=10^{39}\,{\rm erg\,s^{-1}}$, $\varv=2\times 10^3\,{\rm km\,s^{-1}}$ but different p-wind initial densities, and velocities: in the left panel the p-wind velocity is the same as in the i-wind but the density is 60 times higher, while in the right panel, the density is the same as in the i-wind but the velocity is a factor of 6 higher, i.e. $\sim 1.2\times 10^4\,{\rm km\,s^{-1}}$.  The bottom-left panel shows a simulation for twice higher i-wind and p-wind velocities, $4\times 10^3\,{\rm km\,s^{-1}}$  and $\sim 2.4\times 10^4\,{\rm km\,s^{-1}}$, respectively. Finally, the bottom-right panel shows the case when the i-wind is much weaker than in the bottom-left panel. In this case, the jet behavior follows a self-similar solution, when the underdense jet is recollimated by its cocoon. In all simulations, the half-opening angle of the p-wind was the same $\sim 4-5^\circ$}.
\label{fig:jets}
\end{figure*}

To better illustrate the difference between the proposed anisotropic wind model for W50 from a bare jet in a uniform medium, we run four simulations varying i-wind and p-wind parameters (Fig.~\ref{fig:jets}). In all runs the parameters of the ambient medium were the same and the effective half-opening angle $\theta$ of the p-wind was set to $\sim 4^\circ$. This value of $\theta$ also characterizes a smooth transition between the two wind components.  This was done by modulating the angular dependence of the p- and i-winds by $f(\theta)$ and $(1-f(\theta))$, respectively, where $f(\theta)=\cos^n \theta$. Here the value of $n$ controls the confinement angle (see Fig.~\ref{fig:sketch} for the example of a purely conical i-wind). For the simulations shown in Fig.~\ref{fig:jets}, $n=256$ was used and $f(\theta)=0.5$ for $\theta=4.2^\circ$.

Both i- and p-winds are initiated within a $r=8\,{\rm pc}$ sphere.
%with constant radial velocities and balanced pressure.  
The top row compares the cases when the p-wind has the same velocity as the i-wind but 60 times higher density (top-left panel) and the case when the i-wind but the same density but a factor of 6 higher velocity (top-right panel). Broadly, the first case illustrates dense streams of the Haro-Herbig-type, while the second case is more reminiscent of fast extragalactic jets. The kinetic luminosities of the p-wind $L_p\propto \Omega_p \rho_p \varv_p^3$ vary from panel to panel. In the top-left panel, $L_p\sim 0.23\, L_i$, in the top-right and bottom-left panels, $L_p\sim 0.84\,L_i$. For these three panels, $L_i=10^{39}\,{\rm erg\,s^{-1}}$. In the bottom-right panel, the kinetic power of the isotropic wind is very small, while $L_p$ is $0.84\times 10^{39}\,{\rm erg\,s^{-1}}$.

The top-left panel illustrates the "dense jet" case when the p-wind remains strongly overdense with respect to the i-wind even when the latter passes through its termination shock and becomes 4 times denser. In this case, no cocoon is expected to be formed and the recollimation is purely driven by the increased pressure of the i-wind. In this run, the "ears" have relatively simple morphology due to the high momentum of the p-wind.

For the lighter p-wind (top-right panel), the p-wind recollimates itself early on, followed by further recollimation after the i-wind termination shock. After crossing the forward shock this lighter p-wind creates a sequence of eddies in the "ears". 

The bottom-left panel uses a similar setup of a "light" p-wind, except that the velocities of the i- and p-winds are increased by a factor of 2 while keeping the same kinetic luminosities. As expected, the radius of the i-wind terminal shock reduces by a factor $\sim \sqrt{2}$ and so does the recollimation region of the p-wind. Except for these changes, the overall morphology remains qualitatively similar to the top-right panel. 

Finally, the bottom-right panel shows the case when the i-wind is very weak compared to the p-wind, whose parameters are the same as in the bottom-left panel. In this case, the p-wind follows a canonical self-similar pattern expected for underdense jets \citep[e.g.][]{1998MNRAS.297.1087K,2006MNRAS.368.1404A,2012MNRAS.427.3196K}. It develops its own cocoon that recollimates the jet, which then propagates through the uniform medium. The jet keeps a very regular morphology and propagates faster than in other cases. The time of the shown snapshot is $\sim 48\,{\rm kyr}$, i.e. smaller than for other panels  ($\sim 60\,{\rm kyr}$).

This comparison shows that anisotropic winds can broadly reproduce W50 morphology and can feature a recollimation shock inside the forward shock of the i-wind. The complicated substructure of the soft X-ray emission in the lobes of W50  seems to favor lighter/faster p-wind scenarios (see Fig.~\ref{fig:jets}), broadly consistent with simulations of super-Eddington flows \citep[e.g.][]{2022PASJ...74.1378Y}. However, the parameters of the actual outflow from SS433 might be much more complicated, especially if the disk precession is considered, and some of the parameters, like the opening angle of the p-wind are set by hand. While these axisymmetric pure hydrodynamical simulations clearly represent a gross oversimplification of the real system, we believe that they illustrate the main idea of this study.

%-------------------------

\section{Role of SNR}
\label{app:snr}

\begin{figure}
\centering
\includegraphics[angle=0,trim=1cm 5.5cm 1cm 2.5cm,width=0.95\columnwidth]{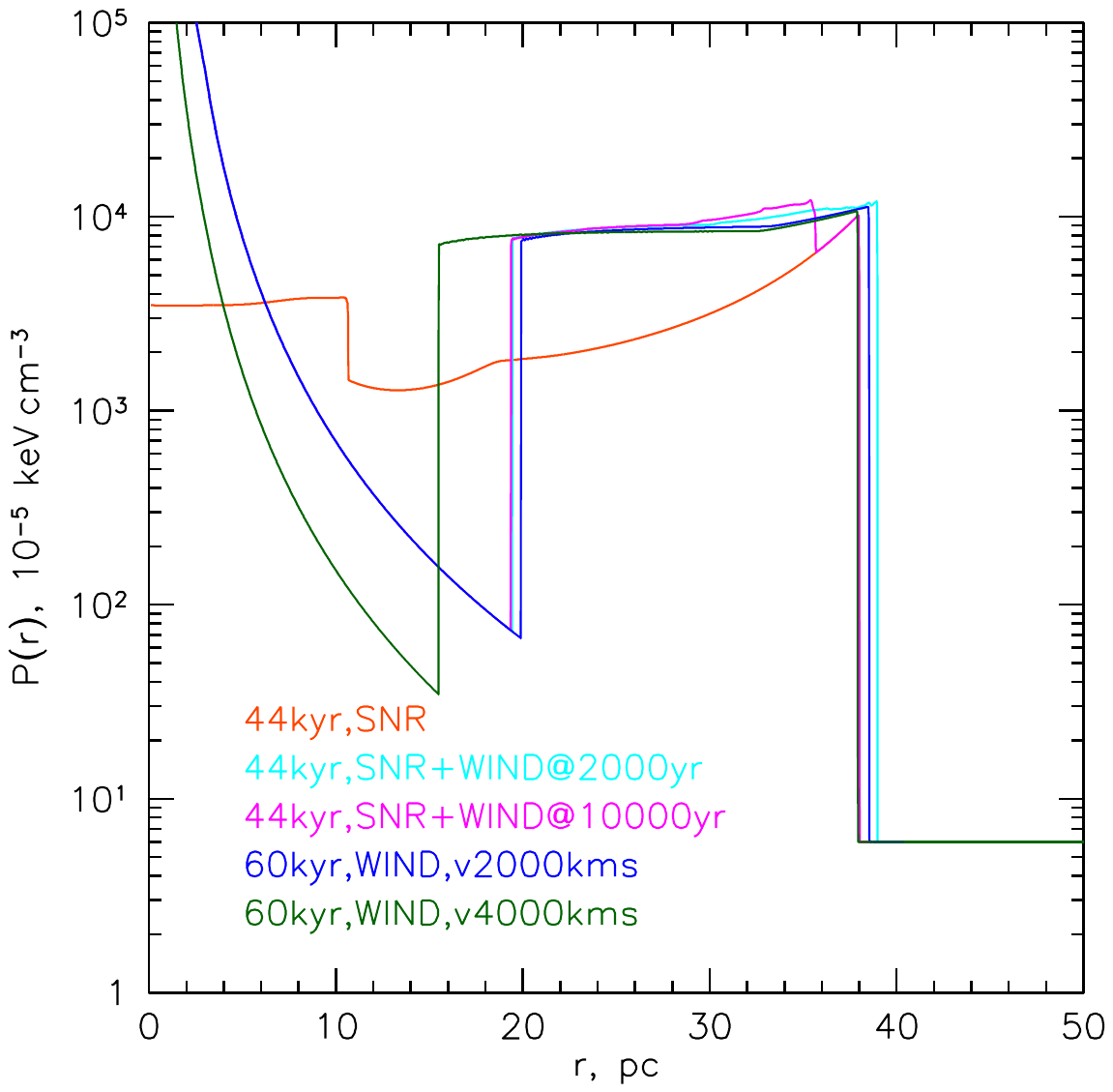}
\caption{Comparison of the gas pressure profiles for cases with and without prior supernova explosion. The blue line shows the case of the isotropic wind operating for 60~kyr in a uniform medium (same as shown in Fig.\ref{fig:wind_def}). The red line shows the pressure profile that would appear if a type II supernova (ejecta mass $20\,M_\odot$, ejecta kinetic energy $10^{51}\,{\rm erg}$) goes off in the same environment. In this case, the forward shock would reach the same radius in $\sim 44\,{\rm kyr}$. The "edges" in the red curve at radii between 10 and 20~pc are the artifacts of the 1D model caused by the reflection of the reserve shock from the center. Finally, the cyan and magenta lines show the pressure profiles when the wind starts operating 2000 and 10,000 yr after the SNR explosion, respectively. In these cases, the profiles shown correspond to $44\,{\rm kyr}$ since the onset of the supernova.  In all cases, a large pressure jump is present at the position of the wind termination shock. The only noticeable effect of the SNR is that the forward shock reaches $r\sim 40\,{\rm pc}$ in a shorter time (44~kyr vs 60~kyr). Finally, the green line shows the case of a twice faster wind (4000 vs 2000 ${\rm km\,s^{-1}})$. As expected, the position of the termination shocks moves closer to the source.
}
\label{fig:snr}
\end{figure}

As discussed in the text, our baseline model assumes that the quasi-spherical part of the W50 nebula is created by the isotropic part of the hyper-accreting compact object on timescale $\sim 60\,{\rm kyr}$ rather than by the supernova explosion that produced of the black hole. In our model, the supernova explosion happened a long time (millions of years) ago. 
However, some scenarios do assume that this supernova is a more recent event $\sim 10^5\, {\rm yr}$ \citep[e.g.][]{2011MNRAS.414.2838G,2017A&A...599A..77P,2018A&A...617A..29B}.  Here we briefly consider a case when an isotropic wind propagates through the still-expanding supernova remnant (SNR). Basically, we consider a type II supernova (ejecta mass $20\,M_\odot$, ejecta kinetic energy $10^{51}\,{\rm erg}$) in a uniform medium and then initiate an isotropic wind with some delays (2000 and 10000~yr) after the SN explosion. Corresponding pressure profiles are shown in Fig.~\ref{fig:snr}. 
All cases are shown at a time when the forward shock reaches $r\sim 38\,{\rm pc}$, which is the observed size of the W50 quasi-spherical part. For comparison, two "pure wind" profiles for different wind velocities are also shown. In all cases, the kinetic power of the isotropic wind is the same ($10^{39}\,{\rm erg\,s^{-1}}$). From this figure, it is clear that in the presence of the SNR, the forward shock region does not change much, although it takes less time compared to the case of the wind alone. The position of the wind termination is not affected, which is the main point of the model considered here.    

%------------------------
%-------------------------

\end{appendix}

%%%%%%%%%%%%%%%%%%%%%%%%%%%%%%%%%%%%%%%%%%%%%%%%%%

% Don't change these lines
%\bsp	% typesetting comment
\label{lastpage}
\end{document}